# AlertMix: A Big Data platform for multi-source streaming data

Ayush Singhal
Contata Solutions
Minneapolis, MN, USA
asinghal@contata.com

Rakesh Pant
Contata Solutions
Minneapolis, MN, USA
rpant@contata.com

Pradeep Sinha
Contata Solutions
Minneapolis, MN, USA
psinha@contata.com

## Abstract

The demand for stream processing is increasing at an unprecedented rate. Big data is no longer limited to processing of big volumes of data. In most real-world scenarios, the need for processing stream data as it comes can only meet the business needs. It is required for trading, fraud detection, system monitoring, product maintenance and of course social media data such as Twitter and YouTube videos. In such cases, a "too late architecture" that focuses on batch processing cannot realize the use cases. In this article, we present an end to end Big data platform called AlertMix for processing multi-source streaming data. Its architecture and how various Big data technologies are utilized are explained in this work. We present the performance of our platform on real live streaming data which is currently handled by the platform.

## Introduction

The term "Big data" is applied to datasets that have size or type which is beyond the ability of traditional relational databases to capture, manage, and process the data with low-latency. As shown in figure 1, Big data is characterized by the 3 V's – high volume, high velocity, or high variety [1]. Such data comes from sensors, devices, video/audio, networks, log files, transactional applications, web, and social media - much of it generated in real time and in a very large scale.

Streaming data is the integral part of most of the business problems with use case spanning from network monitoring[2]–[4] , intelligence and surveillance[5], [6] , e-commerce [7] , fraud detection[8] , smart order routing [9], transaction cost, algorithmic trading, literature services [10]–[15], news and several other real-world use case scenarios. Data from such sources is generated every second or even to the scale of milliseconds. In most such systems what is required is a real-time or near real-time response after processing of the data. Such use case disqualifies the use of Big data architectures such as Hadoop for performing live analysis as the data is generated from the sources. Batch data processing is not a solution for such use cases.

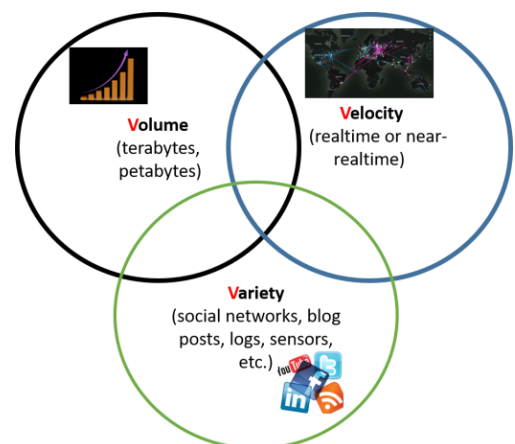

Figure 1: A schematic of 3 V's of Big Data

Streaming data processing platforms are the solution to address such use cases. Till data there are several stream data management platforms are offered such as AWS's Kinesis[1] , IBM's InfoSphere[2], Google's Cloud DataFlow[3], Apache Storm[4] or Spark[5], TIBCO's StreamBase[6]. However, these systems cannot be directly used when the data has to be streamed from thousands of sources with the possibility of adding or deleting older data sources. Such a use case arises in the case of news data collection, processing and delivery to multiple channels in near real time.

In this paper, we discuss an in-house streaming data collection, processing and delivering platform. The proposed platform is based on an actor based system called Akka Streams. In the recent times, Actors systems such as Akka have gained popularity as good platforms for building robust asynchronous data pipelines. Akka Stream, in particular, offers high-level access to design data pipelines. Akka Streams are a streaming interface on top of the Akka actor system.

We present a Akka based platform for processing news feeds from thousands of news RSS feeds. The platform provides a robust approach of selecting data sources, automated data collection from these sources, and real time data processing and multi-channel distribution. In the next section we discuss some background on the Akka Stream system and then detail about out platform construction for news data ingestion and its multi-channel distribution.

The rest of the paper is organized in the following manner. Section 1 provides a brief background of Akka Stream and the actor based model. In Section 2 we discuss the details of the proposed approach. The experiments to validate the performance of the proposed model are described and discussed in Section 3. Finally, Section 4 concludes our work and provides direction for future research.

## Akka Stream

In Akka Streams, the data processing is represented in a form of data flow through an arbitrarily complex graph of processing stages. Stages can have zero or more inputs and zero or more outputs. The pipeline is made of basic building blocks called *Source*s (one output), *Sink*s (one input) and *Flow*s (one input and one output). Using these, arbitrary long linear pipelines can be built.

Akka Streams needs a *Materializer* to execute the pipeline. This materializer is a special tool that actually runs streams, performs necessary and starts all the mechanics. The library in Akka Stream includes an *ActorMaterializer* that executes stream stages on top of Akka actors.

The actor model (proposed by Carl Hewitt) is a way to handle parallel processing in a high-performance network. The actor model overcomes the challenges encountered by a traditional object-oriented programming (OOP) model. The actor model is not only recognized as a highly effective solution but it has been proven in production for some of the world's most demanding applications.

---

[1] https://aws.amazon.com/kinesis
[2] https://www.ibm.com/analytics/information-server
[3] https://cloud.google.com/dataflow/
[4] storm.apache.org/
[5] https://spark.apache.org/
[6] https://www.tibco.com/products/tibco-streambase

## Proposed approach

In this section, we describe the details of the system architecture of the proposed big data solution. The architecture described below is an end to end design of our big data solutions for handling multiple data sources (as shown in figure 2).

**RSS feed collector**

The first component of the system is an RSS feed collector platform. An RSS feed collector is the main platform for gathering data from various streaming feeds. This module allows to connect to multiple stream sources in an incremental manner i.e. news sources can be added or removed on an ongoing basis.

**Bootstrapper**
Bootstrapper will boot up the entire Akka system and will start a scheduler and scheduler's responsibility will be to start Streams picker actor in a pre-configured time interval say every 15 minutes.

**StreamsPickerActor**
Whenever invoked this actor will pick a batch of streams from Couchbase to be processed. Streams will be picked based on their next due date, and also streams which were picked earlier, but could not be updated even after a given time elapsed will also be picked. Picked streams will be updated in couchbase with in-process status. Picked streams will be iterated and passed on to Channel Distributor Actor.

**ChannelDistributorActor**
This actor's responsibility is to find out different channels within the stream and pass those on to appropriate routers for processing. This will also have a bounded priority mailbox.

**PriorityStreamsActor**
This actor will be invoked most likely from AlertMix web application, where by some streams e.g. newly created stream etc. will be processed on priority. It will set priority and pass those stream to channel distributor actor.

**Facebook, Twitter, News, Custom RSS Routers**
Those will be implemented as a balancing pool routers. It will redistribute work from busy routees to idle routees. All routees share the same mail box.

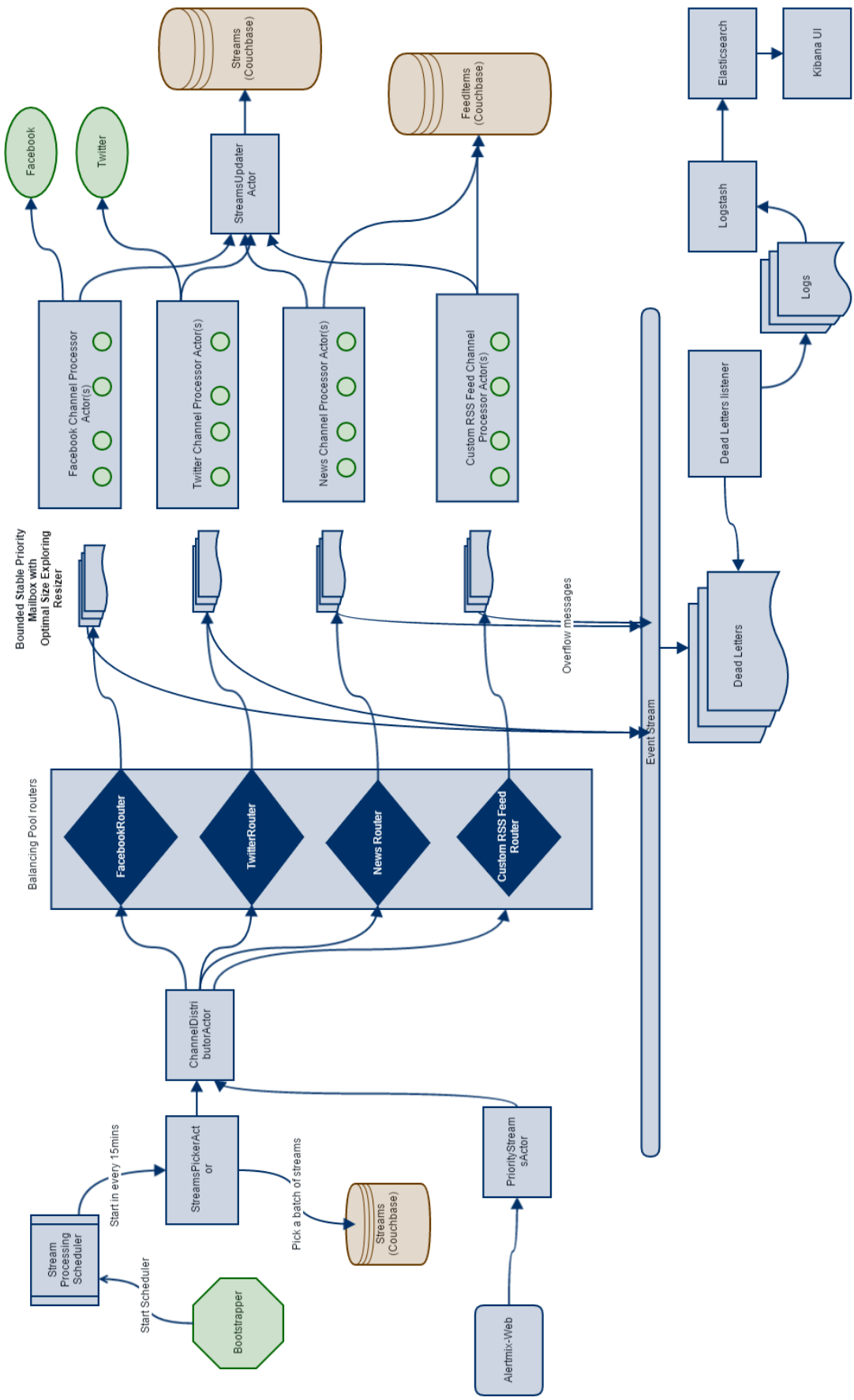

Figure 2: A schematic of the architecture of the proposed model

### Facebook/Twitter and News/Custom RSS Channel Processor Actors

Those will be implemented as a pool of actors with bounded stable priority mail box. Bounded mail box is required to apply back pressure and to avoid long backlog being created which eventually might result in out of memory exception. Priority mail box is required to enable on priority message processing.

### Optimal Size Exploring Resizer

This resizer resizes the pool to an optimal size that provides the most message throughput. Facebook/Twitter actors will call facebook and twitter APIs respectively to get the data. News/Custom RSS actors will get data from couchbase. Its RSS data collector's responsibility to fetch, parse, enrich RSS and news related data.

### StreamsUpdaterActor

Streams updater Actor's responsibility will be to update couchbase with data received for streams and also mark stream's status as processed and update next due date. It will also have a bounded priority mail box.

### DeadLettersListener

As we are using bounded mailboxes at most of the places to apply back pressure, so overflow messages will come to dead letter mail box. This listener will subscribe to dead letters mail box and will generate logs for monitoring purposes and ELK (elasticsearch, logstash, Kibana) stack will be used for monitoring purposes and if it sees unexpected number of dead letters it will email to support group as well.

### Supervisor Strategy

Akka allows us to build system that self-heals and supervisor strategy plays an important role in it. While defining actors we will have to carefully strategize its supervision also, depending on what external system actor is dealing with or what kind of exceptional condition is expected for a given actor. Although, it will be a lower level implementation detail for each actor.

### Message delivery Guarantee

Akka doesn't guarantee delivery of the messages. Although we can ourselves provision for actors to ensure guarantee, but it may become more of an overhead rather than adding value.

Because we have persistent storage of streams, so even if any message is lost and processing of any stream fails it will automatically be picked in next cycles.

In the next section, we discuss the SQS queue pull logic

## SQS Queue Pull Logic

**The main component of a SQS Queue Pull Logic is described as follows:**

**Main SQS Queue** - The main queue for storing Feed messages

**Priority SQS Queue** - The queue which contains newly added Feed messages by users. The messages in this queue are handled with higher priority

**Cron** - Runs at fixed intervals (say 5 seconds), querying the Couchbase[7] database to fetch Feed messages which have their next run time within the next interval (5 seconds)

**FeedRouter** - Supervises the Feed processing workers and pulls messages from the 2 SQS queues with the following logic:
  a. Aims for keeping a certain optimal number of items in the worker-pool mailbox.
  b. As soon as a certain configurable number are processed, uses that as trigger to fetch more items.
  c. Uses a configurable timeout trigger to fetch items from SQS anyway if the configured time has elapsed since the mailbox was last replenished
  d. In both b and c, it tries to replenish the buffer to an optimum size
  e. To handle the scenarios, it programmatically keeps track of the worker mailbox size, last replenishment time and the number of items processed since last replenishment

**Worker** - Receives a feed message, retrieves the feed object from the database and performs a conditional get on the feed based on the eTag and lastModified headers. It handles redirects, checks for duplicate entries already in the system and then processes the results.

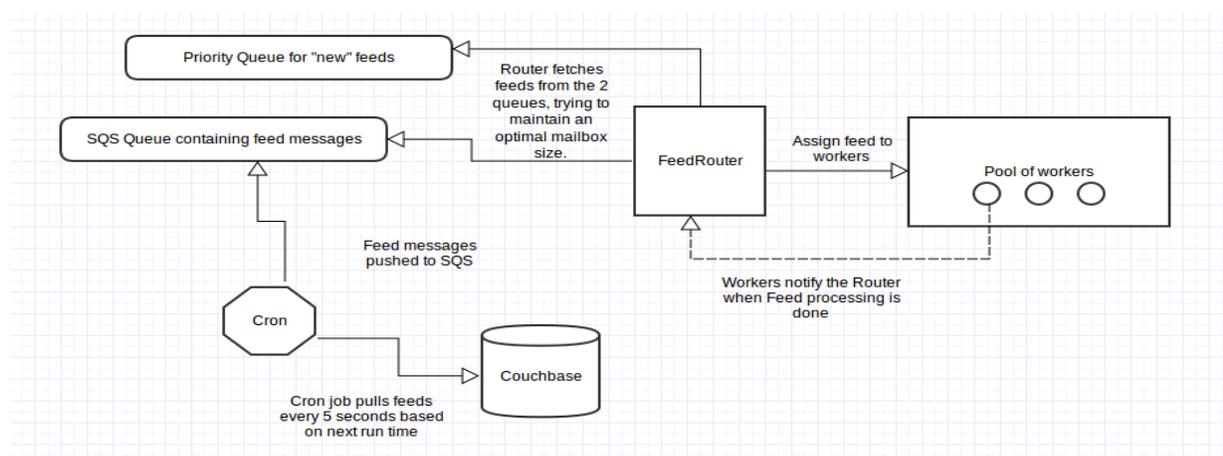

*Figure 3: A schematic of SQS queues.*

# Experiments and Discussion

In this section, we describe the experiment t performed for validating the performance of the proposed model. We use the AWS CloudWatch[8] to monitor the various performance metrics of the proposed model.

---

[7] https://www.couchbase.com/
[8] aws.amazon.com/cloudwatch

We have taken a snapshot of AWS CloudWatch for 24 hours starting from June 17, 2018 3 pm to June 18, 2018 3 pm.

Figure 4 summarizes the performance of the system when processing 200k RSS feeds from various news sources. The system was set to pick up news feeds from these sources every 5 minutes. In the figure, we show the queuing speed of ingestion, sending and deleting. As shown in the figure, the NumberOfMessagesSent chart shows the number of feeds received every 5 minutes. Periodicity trends can be observed in the feeds collected by the system. The peak ingestion in the AWS SQS queue seems to be about 8000 messages in 5 minutes which means about 27 messages/ second.

The proposed platform is able to process and ingest the feeds in the Elasticsearch database maintaining the same queue emptying speed. We find a similar periodicity of queue emptying. This experiment confirms the efficiency of the proposed model to deal with streaming data in real time and avoiding any congestion.

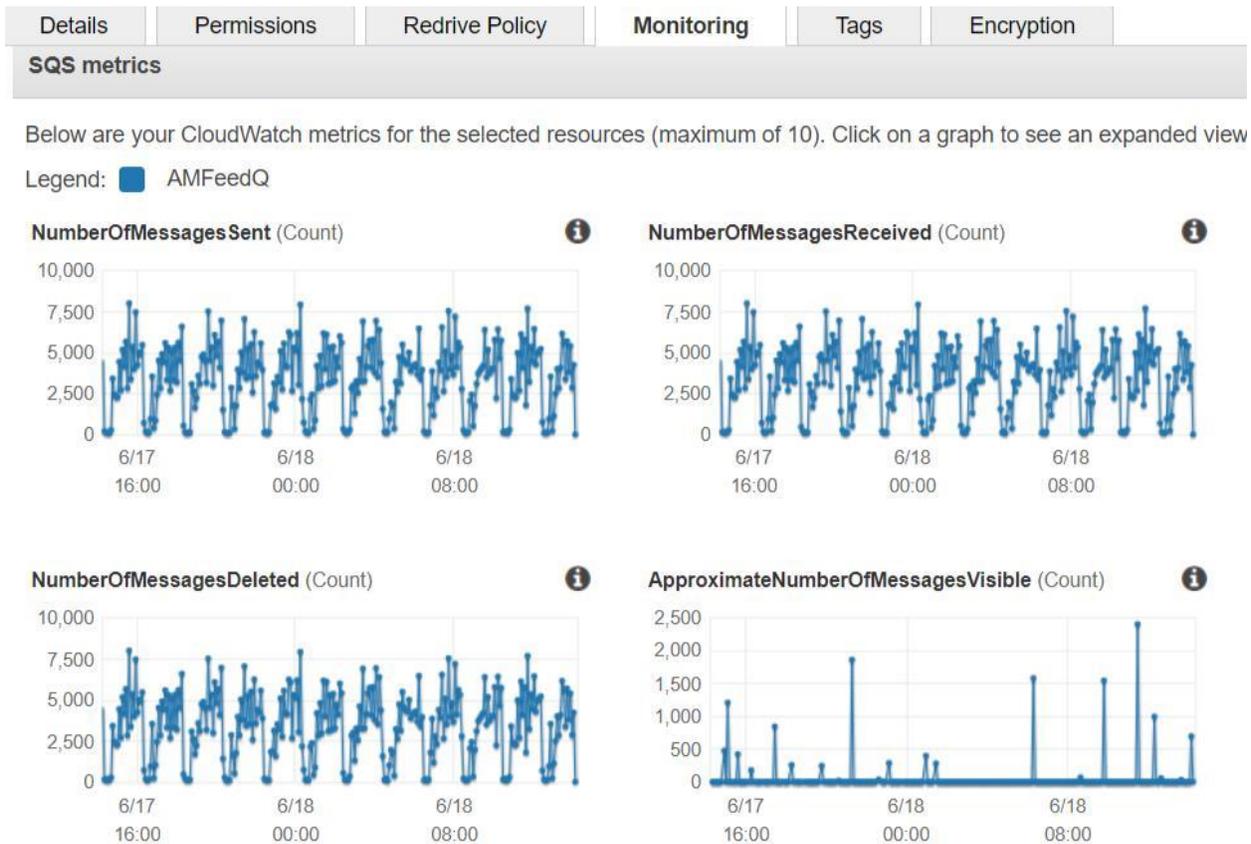

*Figure 4: A screenshot of AWS CloudWatch used for monitoring the performance of the Big Data platform with real load.*

## Conclusions and Future work

In summary, we discussed a Akka based model for processing streaming data from multiple sources. Current industrial solutions provide handling streaming data but such flexibility of adding and removing data sources may be difficult to achieve. The proposed model is ready to deploy solution for any type of streaming data. We presented experiments based on the current working of the system with real data from 200,000 news feed sources and other social media channels such as Facebook and Twitter. The proposed model promises real-time processing of high-speed streaming data.

In the future, the current work can be extended to include more intensive text analytics on the streaming data and still maintaining the real-time efficiency. We also expect to perform several intensive experimentations of the proposed model. We also expect to extend the current service to perform advanced text analytics on Biomedical literature for curating gene-mutation-diseases relationships[16]–[20].